# Predicting the risk of pancreatic cancer with a CT-based ensemble AI algorithm

Chenjie Zhou MD[a, b, c], Jianhua Ma Ph.D [c], Xiaoping Xu MD[a], Lei Feng MD[a], Adilijiang Yimamu MD[a], Xianlong Wang MD [d], Zhiming Li MD [e], Jianhua Mo MS [d], Chengyan Huang MS [d], Dexia Kong MS [d], Yi Gao MD [a, b, *], Shulong Li Ph.D [c,*]

[a] Department of Hepatobiliary Surgery II, Guangdong Provincial Research Center for Artificial Organ and Tissue Engineering, Institute of Regenerative Medicine, ZhuJiang Hospital, Southern Medical University, Guangzhou 510515, Guangdong Province, China.

[b] State Key Laboratory of Organ Failure Research, Southern Medical University, Guangzhou 510515, China

[c] School of Biomedical Engineering, Guangdong Provincial Key Laboratory of Medical Image Processing, Southern Medical University, Guangzhou 510515, Guangdong Province, China

[d] Radiology department, ZhuJiang Hospital, Southern Medical University, Guangzhou 510515, Guangdong Province, China

[e] The Department of radiology, The 2nd affiliated hospital of Guangzhou Medical University, Guangzhou Medical University, Guangzhou 510515, Guangdong Province, China.

Corresponding authors: Yi Gao, MD: drgaoy@126.com;
　　　　　　　　　　Shulong Li, Ph.D: shulong@smu.edu.cn

The word count: 3039

## Abstract

**Objectives:** Pancreatic cancer is a lethal disease, hard to diagnose and usually results in poor prognosis and high mortality. Developing an artificial intelligence (AI) algorithm to accurately and universally predict the early cancer risk of all kinds of pancreatic cancer is extremely important. We propose an ensemble AI algorithm to predict universally cancer risk of all kinds of pancreatic lesions with noncontrast CT.

**Methods:** Our algorithm combines the radiomics method and a support tensor machine (STM) by the evidence reasoning (ER) technique to construct a binary classifier, called RadSTM-ER. RadSTM-ER takes advantage of the handcrafted features used in radiomics and learning features learned automatically by the STM from the CTs for presenting better characteristics of lesions. The patient cohort consisted of 135 patients with pathological diagnosis results where 97 patients had malignant lesions. Twenty-seven patients were randomly selected as independent test samples, and the remaining patients were used in a 5-fold cross validation experiment to confirm the hyperparameters, select optimal handcrafted features and train the model.

**Results:** RadSTM-ER achieved independent test results: an area under the receiver operating characteristic curve of 0.8951, an accuracy of 85.19%, a sensitivity of 88.89%, a specificity of 77.78%, a positive predictive value of 88.89% and a negative predictive value of 77.78%.

**Conclusions:** These results are better than the diagnostic performance of the five experimental radiologists, four conventional AI algorithms, which initially demonstrate the potential of noncontrast CT-based RadSTM-ER in cancer risk prediction for all kinds of pancreatic lesions.

# I. Introduction

Pancreatic cancer is a lethal disease with an incidence that increases yearly, and it is the fourth leading cause of cancer-related deaths worldwide (1). Surgery remains the only curative treatment for pancreatic cancer. However, its prognosis is extremely poor, with a 5-year survival rate of approximately 9% in the United States and 7.2% in China, which is the lowest among ten leading cancer types (2, 3). The deep anatomical location of the pancreas and the biological characteristics of pancreatic cancer make pancreatic cancer able to hide extremely well and difficult to diagnose. Most patients have local invasion and/or distant metastasis at the time of initial diagnosis (2), which seriously affects the prognosis of pancreatic cancer and results in high mortality. Therefore, developing an accurate diagnosis strategy as early as possible is extremely important for improving overall survival, which can benefit patients receiving radical surgery and other precise treatments in a timely manner.

Artificial intelligence (AI) is considered capable of predicting the risk of many diseases, as well as early pancreatic cancer. Such corresponding research has made great progress, especially that based on medical imaging (4). Gulshan et al. first developed a deep learning (DL)-based algorithm for the detection of diabetic retinopathy in retinal fundus photographs (5). A research group from Stanford University also demonstrated an automatic dermatologist-level classifier based on dermoscopic images for identifying different types of skin cancer (6). Titano et al. developed a weakly supervised algorithm with 37,236 head CTs for the surveillance of acute neurologic events (7). Luo et al. developed a real-time AI algorithm for the detection of upper gastrointestinal cancer by endoscopy (8). There is still much successful research on AI in medical applications (9), such as polyp detection in the colon during colonoscopy (10) and prediction of cardiovascular risk factors (11). Most of these studies were modeled based on DL, which requires a large dataset with more than ten thousand cases for training many parameters in DL architectures. Currently, for some clinical problems, such as pancreatic cancer, obtaining such a large dataset is still challenging for the following reasons: (1) the available data from one medical center are limited because of its relatively low incidence; (2) data collection from multiple centers is still hard because the data standardization and sharing problems from different medical centers is still challenging. Therefore, AI algorithm research on smaller datasets is very important and necessary for pancreatic cancer.

AI research in pancreatic cancer has also attracted the attention of many researchers in recent decades, such as malignancy prediction of intraductal papillary mucinous neoplasms (IPMNs) based on quantitative imaging (12) and deep learning (13), computer-aided diagnosis of pancreatic serous cystic neoplasms (PSCNs) based on radiomics methods (14), prediction of the tumor grade based on CT features and texture analysis (15), outcome prediction of stereotactic body radiation therapy (SBRT) based on radiomics signatures (16), prediction of the success of chemotherapy for unresectable pancreatic cancer based on texture analysis (17)

(18) (19), survival prediction based on CT texture analysis (20) and clinical parameters of SEER database (21). However, the research attention on pancreatic cancer is still relatively low. Moreover, most of these studies on the diagnosis or malignancy prediction of pancreatic lesions focused on only one subtype of pancreatic cancer, such as IPMNs (12) (13) and PSCNs (14). One universal AI-aided diagnosis algorithm for all kinds of pancreatic cancer would better satisfy the clinical requirements. Additionally, it would be better to build a model based on a larger dataset with all kinds of data than a model for one subtype. Moreover, there is similarity among all malignant lesions that can be learned with machine learning techniques. Therefore, constructing a universal AI-based diagnosis algorithm for pancreatic cancer has great clinical and technical significance.

The conventional radiomics method with handcrafted features as input has achieved good performance in many medical AI studies (22), especially for tasks with smaller datasets (23, 24) (14). The handcrafted features are predefined by specialists for the special medical task before training the models. However, they are also thought to not reflect the natural disease characteristics (25). Some non-deep machine learning methods can automatically learn natural features from original data, such as support tensor machine (STM), which also have good performance in smaller datasets (26). As an excellent machine learning method for automatically learning features, STM could complement conventional radiomics methods based on handcrafted features. Instead, a radiomics-based algorithm with handcrafted features could also complement an STM-based algorithm. Therefore, combining both methods to construct an ensemble AI algorithm could improve the diagnostic performance for pancreatic cancer.

In particular, for effectively and universally predicting the cancer risk of all kinds of pancreatic lesions, we propose an ensemble AI algorithm using a binary classification method with noncontrast CT. This ensemble AI algorithm combines the conventional radiomics method and the STM method based on the evidence reasoning (ER) technique to construct a binary classifier, called RadSTM-ER. In other words, RadSTM-ER fuses the handcrafted features used in radiomics and learning natural features learned by the STM and can take advantage of both features for presenting better characteristics of lesions.

## II. Materials

### 2.1. Patient cohort

This retrospective study was performed on 194 patients who underwent noncontrast CT scans for initial diagnosis before various treatments and pathological confirmation in the Department of Hepatobiliary Surgery II, Zhujiang Hospital, Southern Medical University, from January 2010 to May 2019. As the gold standard, the diagnostic results with pathology were required to distinguish the benign lesions and malignant lesions for each patient. After removing some cases for missing pathology results, 135 patients were included in this study; 98 patients were diagnosed with malignant lesions and the remaining patients were diagnosed with benign lesions. For all 135 patients, the following detailed characteristics were evaluated: age at the time of treatment, sex, clinical symptoms (including history of pancreatitis), lesion location, imaging findings (including the lesion size and main pancreatic duct [MPD] diameter), pretreatment laboratory values (including carcinoembryonic antigen [CEA], and carbohydrate antigen 19-9 [CA19-9]), family history of pancreatic cancer and whether it was an unexpected discovery or not. For some patients, serum amylase (AMY) and blood lipase (LIP) before treatment were also examined. This study was performed in accordance with the Declaration of Helsinki (27).

**2.2. CT procedure**

For all the patients, CT scans were performed using a Brilliance TM 64 CT or Brilliance TM ICT scanner (Philips Medical Systems Nederland B.V., Best, the Netherlands). The CT volume consisted of different numbers of slices of size 512 ×512 pixels, the same slice thicknesses and reconstruction intervals of 5 mm and different pixel spacings with sizes from 0.5449*0.5449 to 0.9062*0.9062 (mm $^2$).

**2.3. Data preprocessing**

The region of interest (ROI) of each lesion was manually delineated slice by slice by one experimental radiologist, which was further confirmed and amended by another high-level radiologist with 15 years of experience. Subsequently, all 2D ROIs were stacked to 3D ROI for each patient. Because of the pixel spacing difference between different CT scan devices and the large slice thicknesses of 5 mm, the 3D spline interpolation method was performed to fix the imaging resolution to 2 mm*2 mm*2 mm for each lesion.

**2.4. Experimental setup**

As demonstrated in figure 1, we randomly selected 20% of cases (27 cases including 9 benign lesions and 18 malignant lesions) as the independent test subset. For another 80% (108) of cases consisting of 27 benign lesions and 81 malignant lesions, we employed the 5-fold cross validation method to select the optimal feature subset for the radiomics-based method and all hyperparameters in our ensemble algorithm. For one round of 5-fold cross validation experiments, one group of hyperparameters, selected feature subset and the area under the receiver operating characteristic (ROC) curve (AUC) were obtained. The best AUCs were used as the criteria to select the optimal feature subset and hyperparameters. Finally, the ensemble algorithm RadSTM-ER was trained based on all 108 cases with the selected optimal feature subset and hyperparameters, which was tested using the independent subset.

The AUC, accuracy (ACC), sensitivity (SEN), specificity (SPE), positive predictive value (PPV) and negative predictive value (NPV) were used to evaluate the diagnostic performance. The diagnostic performance of four conventional AI methods including the VGG16-based deep learning algorithm, the pure radiomics-based classifier, the pure STM-based classifier and the ensemble algorithm with simple average (RadSTM-Ave), five experimental radiologists were compared with our ensemble algorithms.

## III. Methods

**3.1. The flow of the proposed ensemble algorithm**

The flow of our proposed ensemble AI algorithm is illustrated in figure 2. First, ROIs of the lesions were delineated manually. Second, we extracted the handcrafted features based on the 3D lesion ROIs and constructed the radiomics-based classifier using a support vector machine (SVM) coupled with a forward sequential feature selection method (SFS). Third, we constructed 3D lesion imaging tensors and used an STM-based classifier to automatically learn the natural features (learning features) while performing the classification task. Finally, the reliable analysis-based evidence reasoning (RAbER) method was used to fuse the output scores (probabilities) of both classifiers to obtain the ensemble AI algorithm (RadSTM-ER).

**3.2. The extraction of handcrafted features**

To construct the conventional radiomics-based classifier for the ensemble algorithm, we first extracted 29 high-level handcrafted features (see table 1) consisting of geometric, intensity and texture features extracted from the interpolated 3D lesion ROIs, which have been successfully applied in multiple medical AI tasks (28) (23).

**3.3. The construction of the 3D tensors**

To construct the STM-based classifier, which is an important part of the RadSTM-ER fusion algorithm, 3D lesion tensors of the same size need to be constructed for all the lesions. We used the maximal size of all the lesion ROIs interpolated as the tensor size, which is 71*76*75. For each lesion, a tensor was constructed whose center contained the interpolated 3D ROI and whose periphery zeros were filled. The constructed tensors and their 2D patches in one slice for four cases consisting of two benign lesions and two malignant lesions are shown in figure 3.

**3.4. The radiomics-based classifier**

The conventional radiomics classifier proposed first by Gillies et al. (22) converts the images into mineable data for achieving a particular task, which generally includes three steps: high-throughput extraction of predefined quantitative features, also called radiomics features or handcrafted features, feature selection from these predefined features because of the existing redundancy, and analysis for decision support. In this study, we used SVM coupled with SFS (SVM-SFS) to perform feature selection to obtain the optimal feature subset from the 29 handcrafted features described in subsection 3.2, which were applied successfully in our previous works (29) (23). Subsequently, the selected optimal features were fed into the SVM to perform the classification and obtain the output scores.

**3.5. The STM-based classifier**

We used a kernelled STM (KSTM) (26) with the 3D lesion tensors constructed in subsection 3.3 as input to learn automatically the lesion features, which are also called learning features. Subsequently, the learning features were fed into the SVM to perform the classification and obtain the output scores.

**3.6. The reliable classifier fusion strategy**

We used the reliable analysis based evidence reasoning (RAbER) method (30) (31) to fuse the radiomics-based classifier and the STM-based classifier and obtain the ensemble classification algorithm. This RAbER strategy uses the weight that reflects the relative importance of each fused classifier and the reliability of the output score of each classifier to recalculate the probability scores based on the output scores of both fused classifiers.

**3.7. Statistical analyses**

All statistical analyses were performed using SPSS version 23.0 (SPSS, Chicago, IL). Two-tailed t-tests were used to analyze the significant differences between benign and malignant lesions for all the patient characteristics, and bivariate chi-square tests were used to test the significant differences of the ROCs between our ensemble AI algorithm and the other compared AI algorithms with a P-value <0.05 indicating statistical significance. Logistic regression was used to perform univariable and multivariable analyses.

# IV. Results

**4.1. Characteristics of patients**

The characteristics of the patients are shown in table 2. Ninety-seven patients (71.85%) were pathologically diagnosed with malignancy, which was more than the remaining 38 patients (28.15%) diagnosed with benign lesions. Of the patients with malignant lesions, 55 patients were male; of the patients with benign lesions, 17 patients were male, which indicated that the proportion of malignant lesions in males was higher than that in females. Eight patients with malignant lesions and four patients with benign lesions experienced pancreatitis. Most lesions were located at the pancreatic head, body or tail, and some were located across the head and body, body and tail, and even the head, body and tail. The median and range of CA19-9 and CEA for each patient and AMY and LIP for some patients were accounted for, which showed that all the medians for the patients with malignant lesions were larger than those in the patients with benign lesions. The medians of the lesion sizes and MPD diameter in the patients with malignant lesions are also larger than those in the patients with benign lesions. Five patients had a history of pancreatic cancer in their families, and four patients with benign lesions had a history of pancreatic cancer. Of all patients, five patients with malignant lesions and 12 patients with benign lesions were found unexpectedly.

**4.2. Diagnostic performance of our ensemble AI algorithm**

The detailed results are shown in table 3. Our proposed ensemble AI algorithm RadSTM-ER obtained prediction accuracy, sensitivity, specificity, positive predictive value, negative predictive value, and AUC values of 85.19%, 88.89%, 77.78%, 88.89%, 77.78%, and 0.8951, respectively. These results were better than those of the other four compared AI algorithms. Our model also outperformed the five subspecialty-trained China board-certified radiologists with accuracies of 77.78%, 70.37%, 74.07%, 74.07%, and 74.07%

The ROC curves of five AI algorithms are shown in figure 4 (a), and *P*-values in the bivariate chi-square test between the ROC curves of our ensemble algorithm and the other four compared AI algorithms are shown in table 3, which also indicated that our ensemble algorithm was better than the four compared methods. The comparisons with the performance of the five radiologists is shown in figure 4(b), where the red curve is the ROC curve of our ensemble AI algorithm and the five colored circles indicate the diagnosis results of five radiologists. The black circle is located on the ROC curve, which indicates that the ensemble algorithm RadSTM-ER and the fourth radiologist have similar levels. The other four circles are to the right of and under the ROC curve, which indicates that the ensemble AI algorithm has a higher level than these radiologists.

**4.3. Univariable and multivariable analysis of the diagnostic performance**

The results of the univariable and multivariable analysis between the malignancy of pancreatic lesions and the patient characteristics, the diagnosis results of the 5th radiologist who had the best performance among all radiologists and the output scores of our ensemble AI algorithm are shown in table 4. In the univariable analyses, the lesion size ($\geq$ 40 mm), the diagnosis results of the $5^{th}$ radiologist and the diagnosis results of our ensemble AI algorithm had a significant relationship with the malignancy of the pancreatic lesions. These three associated parameters were further subjected to multivariable logistic regression, which showed that the output scores of our ensemble AI algorithm and diagnosis results of the $5^{th}$ radiologist were the identified factors for the malignancy of pancreatic lesions, while our ensemble AI algorithm had a higher odds ratio of 511.46 (95% confidence interval: 226.10–2049.49, P=0.008).

**4.5. The selected features**

The radiomics-based classifier selected six optimal features (table 5), which consisted of three intensity features, two geometric features and a texture feature. The three intensity features are the maximum, minimum and standard deviation of intensity. The two geometric features are minor diameter and the eccentricity of the lesion. The texture feature is the homogeneity.

# V. Conclusions and discussions

In this study, we predicted the universal cancer risk of all kinds of pancreatic lesions using an ensemble AI algorithm. This ensemble AI algorithm combined the radiomics method and an STM to construct a binary classifier using the reliable classifier fusion strategy-based evidence reasoning technique (RadSTM-ER). Thus, the ensemble AI algorithm took advantage of the handcrafted features and the natural learning features from STM of noncontrast CT.

This study did not focus on one pancreatic lesion subtype, such as PSCNs or IPMNs; rather, it universally predicted the malignancy of all kinds of pancreatic lesions and achieved the test performance with a prediction accuracy of 85.19%, a sensitivity of 88.89%, a specificity of 77.78%, a PPV of 88.89%, an NPV of 77.78%, and an AUC of 0.8951. These prediction results were better than the diagnostic performance of five experimental radiologists, four other AI algorithms (see table 3 and figure 4), which indicates the feasibility of the universal AI algorithm.

Because of the weak visual performance of pancreatic disease in noncontrast CT in humans, most radiologists usually diagnose pancreatic disease by combining contrast CT or other clinical parameters. Thus, the diagnostic performance of radiologists based on noncontrast CT does not always achieve the expected result. The performance of our proposed ensemble algorithm achieves a better performance than the radiologists. However, the fusion algorithm combining further contrast CT, serum examination parameters, and symptom information needs to be studied further when the data are complete.

In total, 135 patients were included in this study, comprising 38 patients with benign lesions and 97 patients with malignant lesions. For most medical problems solved by the binary classification method, the majority classes are usually positive classes (malignant cases) (23, 26, 32). However, the majority class is the negative class (benign cases) in this work. Such an opposite phenomenon might be caused by the fact that most patients with a primary diagnosis in the Department of Hepatobiliary Surgery II usually have uncomfortable symptoms and thus have a higher probability of malignancy.

Additionally, because of the small dataset size (135 cases), we combined the handcrafted feature-based radiomics method and the nondeep machine learning method STM, which can automatically learn natural features to construct the ensemble AI algorithm without considering deep learning, although it has much potential in many medical tasks. If more data from different medical centers and institutes are collected, deep learning should be included in the ensemble algorithm to achieve better performance.

**Acknowledgments**

This work was partly supported by the National Natural Science Foundation of China (NSFC, 11771456).

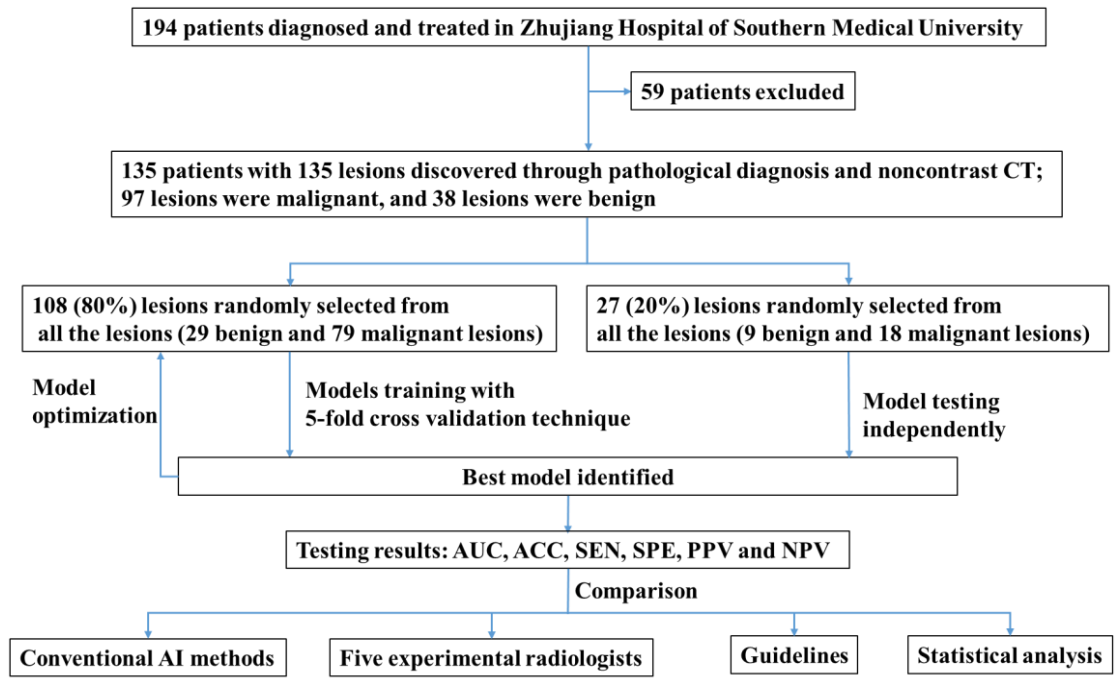

Figure 1. The workflow of the experimental setup. AUC: the area under the ROC curve, ACC: accuracy, SEN: sensitivity, SPE: specificity, PPV: positive predictive value, NPV: negative predictive value.

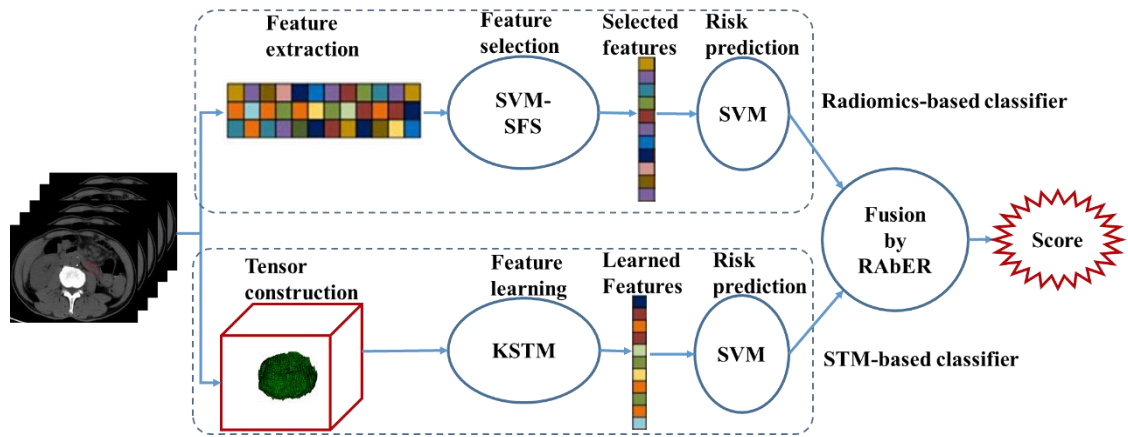

Figure 2. The flow of our proposed ensemble AI algorithm RadSTM-ER. SVM: support vector machine; SVM-SFS: SVM coupled with the forward sequential feature selection method (SFS); KSTM: kernelled support tensor machine; RAbER: the reliable analysis-based evidence reasoning method

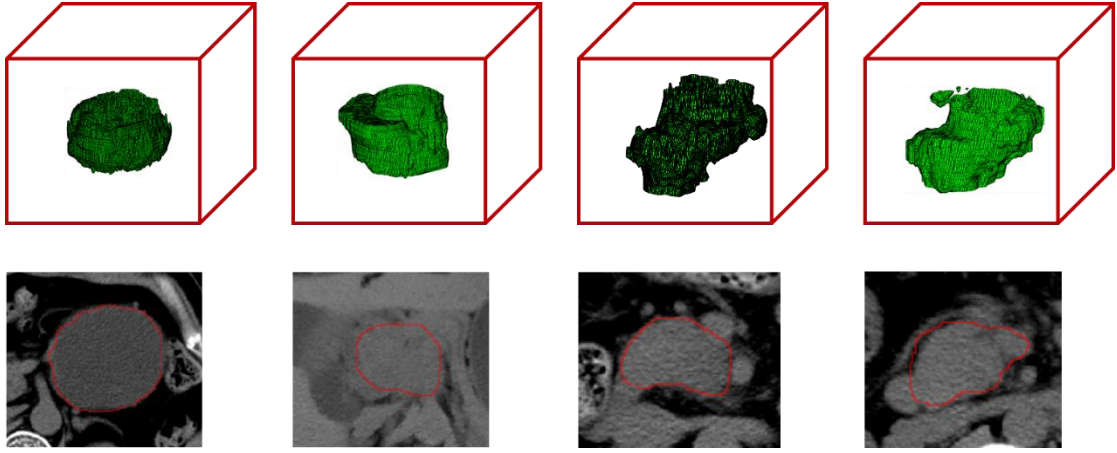

Figure 3. The constructed 3D lesion tensors (top row) and their corresponding 2D patches, including the lesion in one slice (bottom row): the left columns are the benign lesions, and the right columns are the malignant lesions.

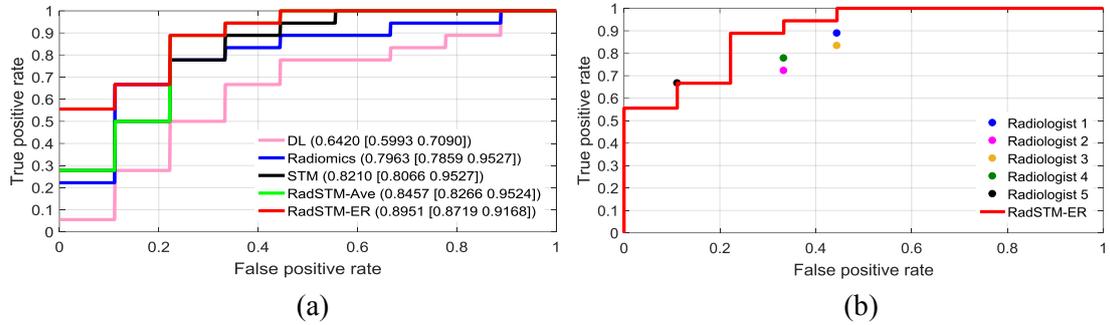

(a)                                      (b)

Figure 3. (a) The ROC curves of the five compared algorithms for the independent test samples; (b) the test performance of our ensemble AI algorithm and the five radiologists. The ROC curve for ensemble AI algorithm (red line) and the radiologist's performance (colored circles) for the diagnosis results with just noncontrast CT.

Table 1. The radiomics features used in our fusion algorithms.

| Intensity features | Geometric features | Texture features |
|---|---|---|
| Minimum | Volume | Energy |
| Maximum | Major diameter | Entropy |
| Mean | Minor diameter | Correlation |
| Standard deviation | Eccentricity | Contrast |
| Sum | Elongation | Texture variance |
| Median | Orientation | Sum-Mean |
| Skewness | Bounding box volume | Inertia |
| Kurtosis | Perimeter | Cluster shade |
| Variance | | Cluster prominence |
| | | Homogeneity |
| | | Max probability |
| | | Inverse variance |

Table 2. Patient characteristics.

| Characteristics | Benign (38) | Malignant (97) | P-value |
|---|---|---|---|
| Age, yrs, median (range) | 50 [18-68] | 60 [26-85] | <0.0001 |
| Male sex, n (%) | 17 (44.74) | 55 (56.70) | <0.0001 |
| History of pancreatitis, n (%) | 4 (10.53) | 8 (8.42) | 0.2391 |
| Lesion location, n (%) | | | -- |
|   Head | 17 (77.74) | 66 (68.04) | |
|   Body | 11 (28.95) | 21 (21.65) | |
|   Tail | 18 (47.37) | 29 (29.90) | |
|   Head/body | 2 (5.26) | 2 (2.06) | |
|   Body/tail | 7 (18.42) | 15 (15.46) | |
|   Head/body/tail | 1 (2.63) | 1 (1.03) | |
| Lesion size, mm, median (range) | 22.02[6.83-60.97] | 25.08 [6.86-99.61] | 0.0254 |
| MPD diameter, mm, median (range) | 2.10 [0.80-7] | 2.60 [0.70-18] | 0.0078 |
| CA 19-9, U/ml, median (range) | 12 [0.50-716.10] | 242.13 [0.50-486930] | 0.3278 |
| CEA, ng/ml, median (range) | 1.50 [0.20-64.50] | 4.00 [0.30-1109] | 0.1697 |
| Family history of PC, n (%) | 4 (10.53) | 5 (5.15) | -- |
| Unexpected discovery, n (%) | 12 (31.58) | 5 (5.15) | -- |
| Serum AMY, IU/L, median (range) | 69.0 [45-126.40] | 71.10 [20.80-415.50] | 0.0241 |
| Serum LIP, IU/L, median (range) | 44.6 [22.7-214.5] | 78.3 [6-1118.7] | 0.0018 |

*Note.* MPD, main pancreatic duct; CEA, carcinoembryonic antigen; CA19-9, carbohydrate antigen 19-9; AMY, serum amylase; LIP, blood lipase

Table 3. The test results predicted by our proposed ensemble algorithm, four compared AI algorithms, five radiologists and four factors with cutoff points reported in guidelines.

| | Methods | ACC (%) | SEN (%) | SPE (%) | PPV (%) | NPV (%) | AUC | P-value |
|---|---|---|---|---|---|---|---|---|
| AI algorithms | RadSTM-ER | 85.19 | 88.89 | 77.78 | 88.89 | 77.78 | 0.8951 | -- |
| | RadSTM-Ave | 81.48 | 88.89 | 66.67 | 84.21 | 75.00 | 0.8457 | <0.000 |
| | Radiomics | 77.78 | 77.78 | 77.78 | 77.78 | 83.33 | 0.7963 | <0.000 |
| | STM | 81.48 | 94.44 | 55.56 | 80.95 | 63.64 | 0.8210 | <0.000 |
| | DL | 62.96 | 66.67 | 55.56 | 75.00 | 45.45 | 0.6420 | <0.000 |
| Radiologists | Radiologist 1 | 77.78 | 88.89 | 55.56 | 80.00 | 71.43 | -- | -- |
| | Radiologist 2 | 70.37 | 72.22 | 66.67 | 81.25 | 54.55 | -- | -- |
| | Radiologist 3 | 74.07 | 83.33 | 55.56 | 78.95 | 62.50 | -- | -- |
| | Radiologist 4 | 74.07 | 66.67 | 88.89 | 92.31 | 57.14 | -- | -- |
| | Radiologist 5 | 74.07 | 77.78 | 66.67 | 82.35 | 60.00 | -- | -- |

Table 4. Univariable/multivariable logistic regression analysis of the relationship between malignancy of pancreatic lesions and other factors.

|  | Univariable | | Multivariable | |
|---|---|---|---|---|
|  | OR (95% CI) | P-value | OR (95% CI) | P-value |
| Age (≥ 70 yrs) | 1.03 [0.98-1.09] | 0.262 | | |
| Sex (male) | 2.00 [0.38-10.58] | 0.415 | | |
| Pancreatitis (+) | 2.29 [0.22-24.14] | 0.492 | | |
| Lesion size (≥ 40 mm) | 0.90 [0.82-1.00] | 0.039 | 0.80 [0.58-1.09] | 0.158 |
| MPD diameter (≥ 10 mm) | 1.57 [0.90-2.76] | 0.115 | | |
| CA 19-9 (≥ 38 U/mL) | 1.00 [0.99-1.01] | 0.209 | | |
| CEA (≥ 5.1 ng/mL) | 1.68 [0.83-3.41] | 0.150 | | |
| Family history (+) | 0.21 [0.02-2.66] | 0.226 | | |
| Radiologist 5 (+) | 16.00 [1.61-159.31] | 0.018 | 53.70 [10.35-81.94] | 0.012 |
| RadSTM-ER (≥ 0.5) | 396.05 [107.14-468.08] | 0.006 | 511.46 [226.10-2049.49] | 0.008 |

*Note.* MPD, main pancreatic duct; CEA, carcinoembryonic antigen; CA19-9, carbohydrate antigen 19-9; AMY, serum amylase; LIP, blood lipase

Table 5. The optimal features selected by the ensemble algorithm

| Intensity features | Geometric features | Texture features |
|---|---|---|
| Maximum | Minor diameter | Homogeneity |
| Minimum | Eccentricity | |
| Standard deviation | | |